\documentstyle[12pt,epsf,epsfig,wrapfig]{article}
\def\beq{\begin{eqnarray}} \def\eeq{\end{eqnarray}}
\newcommand{\beqe}{\begin{equation}} \newcommand{\eeqe}{\end{equation}}

\begin{document}

\begin{center}
{\bfseries SPIN POLARIZATION PHENOMENA IN  DENSE NUCLEAR MATTER }

\vskip 5mm A.A. Isayev

\vskip 5mm {\small  {\it Kharkov Institute of Physics and
Technology,
 Kharkov, 61108, Ukraine
}}
\end{center}

\vskip 5mm
\begin{abstract}
Spin polarized states in nuclear matter with  an effective
nucleon-nucleon interaction are studied for a wide range of
isospin asymmetries and densities. Based on a Fermi liquid theory,
it is shown that there are a few possible scenarios of spin
ordered phase transitions: (a) nuclear matter undergoes at some
critical density a phase transition to a spin polarized state with
the oppositely directed spins of neutrons and protons (Skyrme SLy4
and Gogny D1S interactions); (b) at some critical density, a spin
polarized state with the like-directed neutron and proton spins
appears (Skyrme SkI5 interaction); (c) nuclear matter under
increasing density, at first, undergoes a phase transition to the
state with the opposite directions of neutron and proton spins,
which goes over at larger density to the state with the same
direction of nucleon spins (Skyrme SkI3 interaction).
\end{abstract}

\vskip 8mm

The issue of spontaneous appearance of  spin polarized states in
nuclear matter is a topic of a great current interest due to its
relevance in astrophysics. In particular, the scenarios of
supernova explosion and cooling of neutron stars are essentially
different, depending on whether nuclear matter is spin polarized
or not. On the one hand, the models with the effective
nucleon-nucleon (NN) interaction predict the occurrence of spin
instability in nuclear matter at densities in the range from
$\varrho_0$ to $6\varrho_0$ for different parametrizations of the
NN potential~\cite{KW94}--\cite{IY}
($\varrho_0=0.16\,\mbox{fm}^{-3}$). On the other hand, for the
models with the realistic NN interaction, the ferromagnetic  phase
transition seems to be suppressed up to densities well above
$\varrho_0$~\cite{PGS}--\cite{FSS}.

 Here the issue  of spin polarizability of nuclear matter is considered with the use
of an effective NN interaction. The main objective is to study the
possible scenarios of spin ordered phase transitions in dense
nuclear matter with Skyrme and Gogny forces.   In particular, we
choose Skyrme SLy4  effective interaction, constructed originally
to reproduce the results of microscopic neutron matter
calculations~\cite{CBH}. We utilize Skyrme SkI3 and SkI5
parametrizations as well, giving a correct description of isotope
shifts in neutron-rich medium and heavy nuclei~\cite{RF}. Besides,
we employ Gogny D1S interaction, widely used in nuclear structure
calculations. The basic formalism is presented in detail in
Ref.~\cite{IY}. We are interested in studying   spin polarized
states with like-directed and oppositely directed spins of
neutrons and protons. One should solve the self-consistent
equations for the coefficients
$\xi_{00},\xi_{30},\xi_{03},\xi_{33}$ in the expansion of the
single particle energy in Pauli matrices in spin and isospin
spaces \beq\xi_{00}({\bf p})&=&\varepsilon_{0}({\bf
p})+\tilde\varepsilon_{00}({\bf p})-\mu_{00},\;
\xi_{30}({\bf p})=\tilde\varepsilon_{30}({\bf p}),\label{14.2} \\
\xi_{03}({\bf p})&=&\tilde\varepsilon_{03}({\bf p})-\mu_{03}, \;
\xi_{33}({\bf p})=\tilde\varepsilon_{33}({\bf p}).\nonumber\eeq
Here $\varepsilon_0({\bf p})$ is the free single particle
spectrum, $\mu_{00}$ and $\mu_{03}$ are half of a sum and half of
a difference of neutron and proton chemical potentials,
respectively, and
$\tilde\varepsilon_{00},\tilde\varepsilon_{30},\tilde\varepsilon_{03},\tilde\varepsilon_{33}$
are the Fermi liquid (FL) corrections to the free single particle
spectrum, related to the normal FL amplitudes $U_0({\bf
k}),...,U_3({\bf k}) $ by formulas \beq\tilde\varepsilon_{00}({\bf
p})&=&\frac{1}{2\cal V}\sum_{\bf q}U_0({\bf k})f_{00}({\bf q}),\;
 \tilde\varepsilon_{30}({\bf p})=\frac{1}{2\cal V}\sum_{\bf
q}U_1({\bf k})f_{30}({\bf q}),\; {\bf k}=\frac{{\bf p}-{\bf q}}{2},
\label{14.1}\\ 
\tilde\varepsilon_{03}({\bf p})&=&\frac{1}{2\cal V}\sum_{\bf
q}U_2({\bf k})f_{03}({\bf q}), \;\tilde\varepsilon_{33}({\bf
p})=\frac{1}{2\cal V}\sum_{\bf q}U_3({\bf k})f_{33}({\bf q}).
\nonumber \eeq The distribution functions
$f_{00},f_{03},f_{30},f_{33}$, in turn, can be expressed in terms
of  the components $\xi$ of the single particle energy and satisfy
the normalization conditions for the total density
$\varrho_n+\varrho_p=\varrho$, excess of neutrons over protons
$\varrho_n-\varrho_p\equiv\alpha\varrho$, ferromagnetic (FM)
$\varrho_\uparrow-\varrho_\downarrow\equiv\Delta\varrho_{\uparrow\uparrow}$
and antiferromagnetic (AFM)
$(\varrho_{n\uparrow}+\varrho_{p\downarrow})-
(\varrho_{n\downarrow}+\varrho_{p\uparrow})\equiv\Delta\varrho_{\uparrow\downarrow}$
spin order parameters, respectively  ($\alpha$ being the isospin
asymmetry parameter,
$\varrho_\uparrow=\varrho_{n\uparrow}+\varrho_{p\uparrow}$ and
$\varrho_\downarrow=\varrho_{n\downarrow}+\varrho_{p\downarrow}$,
with  $\varrho_{n\uparrow},\varrho_{n\downarrow}$
 and
 $\varrho_{p\uparrow},\varrho_{p\downarrow}$ being the neutron and
 proton number densities with spin up and spin down).
  The quantities of interest are  the neutron and
proton spin polarization parameters
$\Pi_n=\frac{\varrho_{n\uparrow}-\varrho_{n\downarrow}}{\varrho_n},
\Pi_p=\frac{\varrho_{p\uparrow}-\varrho_{p\downarrow}}{\varrho_p},
$
characterizing spin ordering in neutron and proton subsystems.
\begin{figure}[tb]
\begin{center}
\mbox{\epsfig{figure=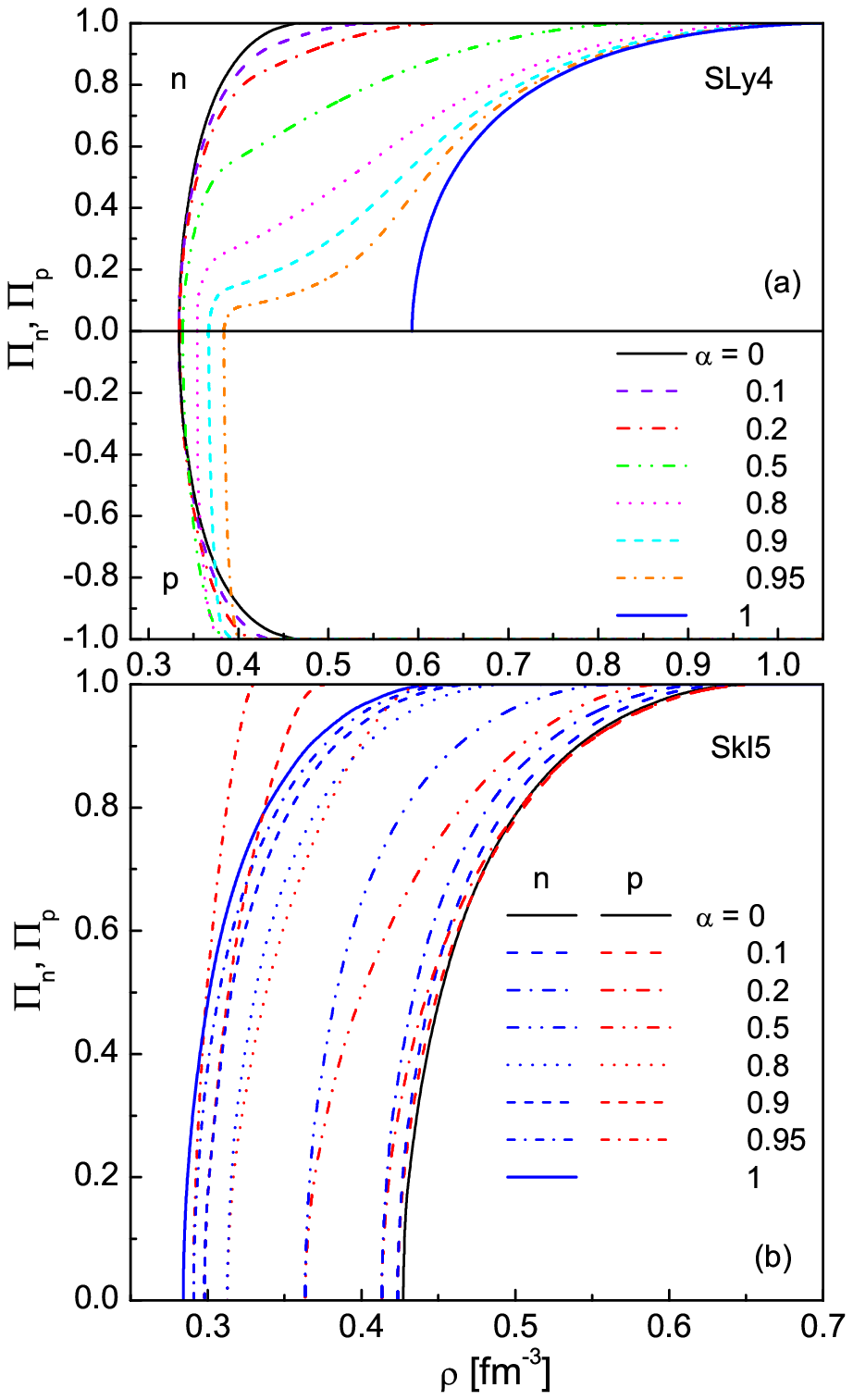,height=13.8cm,width=8.6cm,trim=49mm
86mm 56mm 46mm}}
\end{center}
{\small{\bf Figure~1.} Neutron  and proton  spin polarization
parameters as functions of density  at zero temperature for (a)
SLy4 force and (b) SkI5 force.}
\label{fig1}
\end{figure}

Fig.~1a shows the density dependence of the neutron and proton
spin polarization parameters at zero temperature for SLy4 force.
The main qualitative feature is that for SLy4 force there are only
solutions corresponding to the oppositely directed spins of
neutrons and protons in a spin polarized state. The reason is that
for SLy4 force the FL amplitude $U_1$, determining spin--spin
correlations, is repulsive for all relevant densities, while the
 FL amplitude $U_3$, describing spin--isospin correlations, becomes quite
 attractive at high densities.
  The critical
density of spin instability in symmetric nuclear matter
($\alpha=0$), corresponding to AFM spin ordering
($\Delta\varrho_{\uparrow\downarrow}\not=0$,
$\Delta\varrho_{\uparrow\uparrow}=0$), is $\varrho_c\approx0.33$
fm$^{-3}$. It is less than the critical density of FM instability
 in neutron matter, $\varrho_c\approx0.59$ fm$^{-3}$. Even small
admixture of protons to neutron matter leads to the appearance of
long tails in the density profiles of the neutron spin
polarization parameter near the transition point to a spin ordered
state. As a consequence, a spin polarized state is formed much
earlier in density than in pure neutron matter.

As seen from Fig.~1b, for SkI5 force, oppositely to SLy4 force,
there are only solutions corresponding to the same direction of
neutron and proton spins in a polarized state. In the case under
consideration the FL amplitude $U_3$ is repulsive for all relevant
densities, while the
 FL amplitude $U_1$ becomes quite attractive  at high densities. For SkI5 force,
a phase transition to the  FM spin  state in neutron matter takes
place at the critical density $\varrho_c\approx0.28$ fm$^{-3}$. It
is less than the critical density of spin instability in symmetric
nuclear matter $\varrho_c\approx0.43$ fm$^{-3}$, corresponding to
 FM spin ordering ($\Delta\varrho_{\uparrow\uparrow}\not=0$,
$\Delta\varrho_{\uparrow\downarrow}=0$). There are no long tails
in the density profiles of the neutron spin polarization parameter
at large isospin asymmetry. In the given case, a small admixture
of protons to neutron matter even leads  to the increase of the
critical density of spin instability.

\begin{figure}[tb]
\begin{center}
\mbox{\epsfig{figure=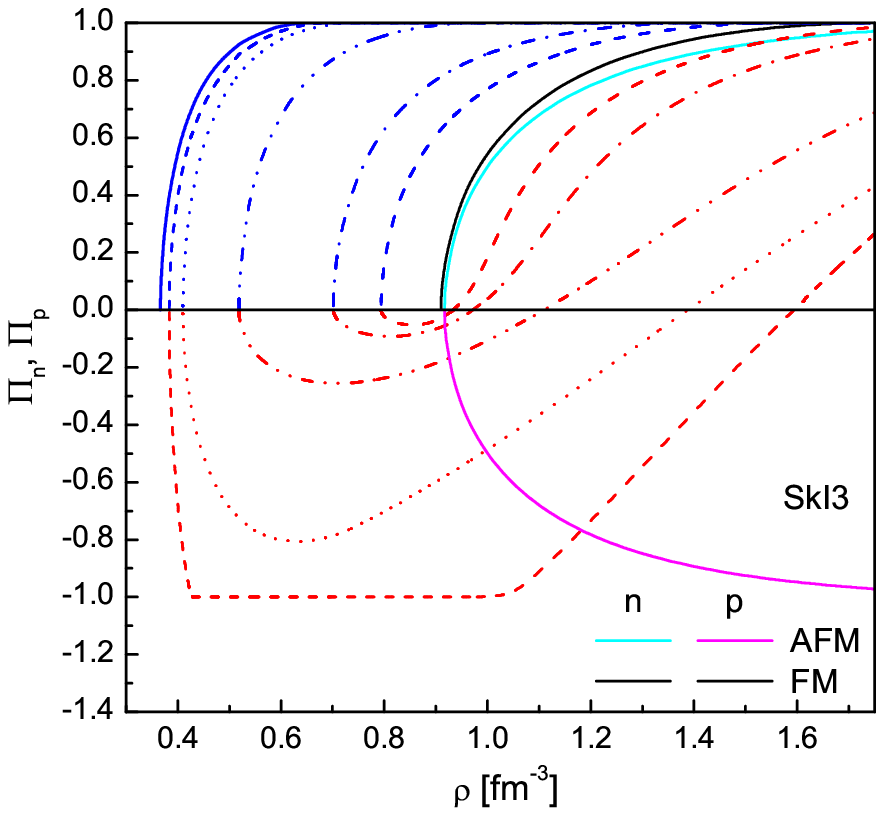,height=7.5cm,width=8.6cm,trim=45mm
134mm 57mm 66mm}}
\end{center}
{\small{\bf Figure~2.}  Same as in Fig.~1, but for SkI3 force.
Also the curves, corresponding to FM and AFM ordering in symmetric
nuclear matter, are shown.} \label{fig2}
\end{figure}

Fig.~2 shows the neutron and proton spin polarization parameters
as functions of density at zero temperature for SkI3 force.  There
are two types of solutions of the self-consistent equations in
symmetric nuclear matter, corresponding to FM and AFM ordering of
neutron and proton spins. Due to proximity of FL amplitudes $U_1$
and $U_3$, the respective critical densities are very close to
each other ($\varrho_c\approx 0.910\,\mbox{fm}^{-3}$ for FM
ordering and $\varrho_c\approx 0.917\,\mbox{fm}^{-3}$  for AFM
ordering) and larger than the critical density of spin instability
in neutron matter ($\varrho_c\approx 0.37\,\mbox{fm}^{-3}$). When
some admixture of protons is added to neutron matter, the last
critical density is shifted to larger densities and a spin
polarized state with the oppositely directed spins of neutrons and
protons appears. Under increasing density of nuclear matter, the
neutron spin polarization continuously increases till all neutron
spins will be aligned in the same direction. Protons, at first,
become more polarized with density and their spin polarization is
opposite to the spin polarization of neutrons. But, after reaching
the maximum, spin polarization of protons decreases and at some
critical density spins of protons change direction, so that the
spin ordered phase with the like-directed spins of neutrons and
protons is formed. Then, beyond the critical density, the spin
polarization of protons is continuing to increase until the
totally polarized state with parallel ordering of neutron and
proton spins will be formed. Thus, for SkI3 force  nuclear matter
undergoes at some critical density a phase transition from the
state with antiparallel ordering of neutron and proton spins to
the state with  parallel ordering of spins. With increasing
isospin asymmetry, this critical density increases as well. Note
that there are no long tails in the density profiles of the
neutron spin polarization parameter at large  asymmetries.

\begin{figure}[tb]
\begin{center}
\mbox{\epsfig{figure=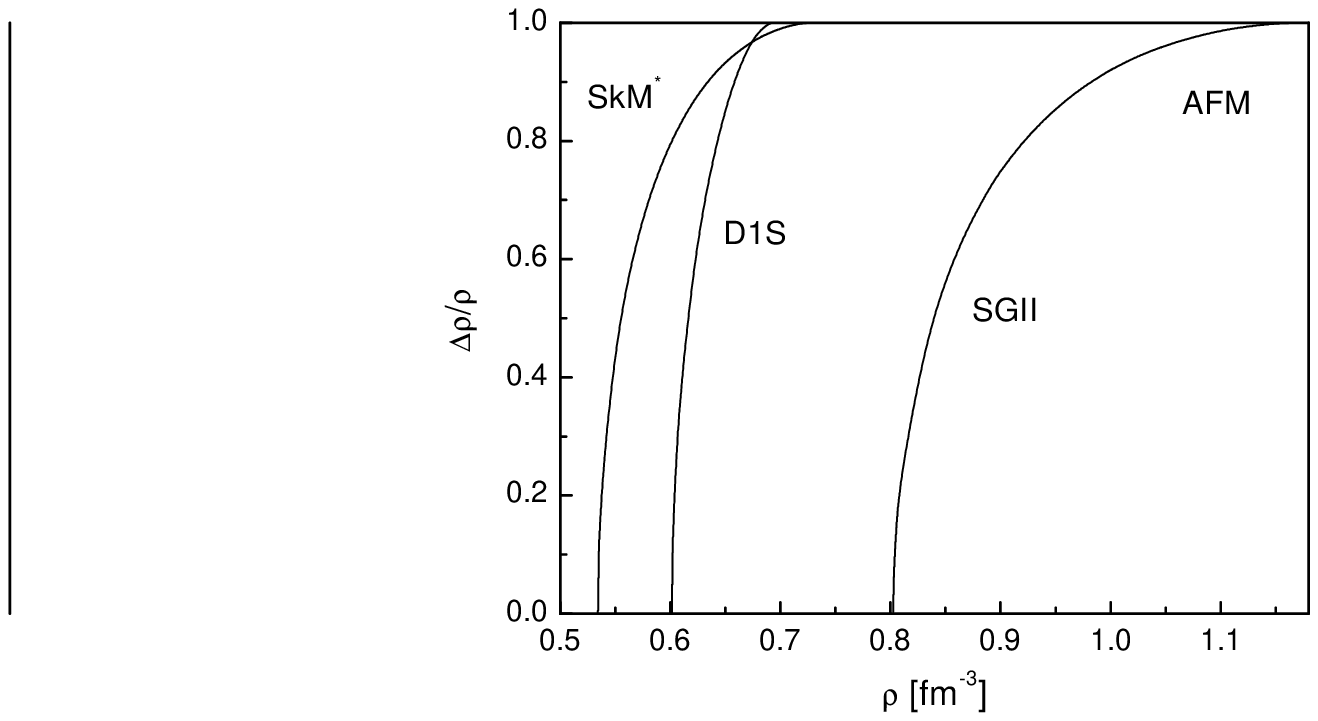,height=7.0cm,width=8.6cm,trim=48mm
142mm 57mm 68mm,clip}}
\end{center}
{\small{\bf Figure~3.}  AFM spin polarization parameter as a
function of density at zero temperature for the D1S
 Gogny force and  the SkM$^*$, SGII Skyrme forces.} \label{fig3}
\end{figure}

Now we present the results of the numerical  solution of the
self--consistent equations with the  D1S  Gogny effective force
for symmetric nuclear matter. The main qualitative feature is that
for the D1S force there are only solutions corresponding to  AFM
spin ordering and there are no solutions corresponding to  FM spin
ordering. In Fig.~3, it is shown the dependence of the  AFM spin
polarization parameter
$\Delta\varrho_{\uparrow\downarrow}/\varrho$  as a function of
density at zero temperature. The AFM spin order parameter arises
at density $\varrho\approx3.8\varrho_0$ for the D1S potential. A
totally antiferromagnetically polarized state
($\Delta\varrho_{\uparrow\downarrow}/\varrho=1$) is formed at
$\varrho\approx4.3\varrho_0$. The neutron and proton spin
polarization parameters for the AFM spin ordered state are
opposite in sign and equal to
$$\Pi_n=-\Pi_p=\frac{\Delta\varrho_{\uparrow\downarrow}}{\varrho}.$$
For comparison, we plot in Fig.~3 the density dependence of the
AFM spin polarization parameter  for the Skyrme effective forces
SkM$^*$ and SGII being relevant for calculations at small isospin
asymmetry.

It is necessary to emphasize that  different behavior at high
densities of the interaction amplitudes, describing spin--spin and
spin--isospin correlations, lays behind this divergence in
calculations with different effective forces. These results
clearly indicate the necessity to construct a new generation of
the energy functionals with the properly constrained time-odd part
at high densities. Probably, these constraints will be obtained
from the data on the time decay of magnetic field of isolated
neutron stars~\cite{PP}. \eject

\end{document}